%% file: paper.tex
\documentclass[sigconf]{acmart}

\newif\ifAPPENDIX
\APPENDIXtrue
\APPENDIXfalse

\copyrightyear{2026} 
\acmYear{2026} 
\setcopyright{rightsretained} 
\acmConference[ICSE '26]{48th IEEE/ACM International Conference on Software Engineering}{2026}{Rio De Janeiro, Brazil}
\acmDOI{DOI} 
\acmISBN{ISBN}

\newif\ifDEBUG
\DEBUGfalse
\DEBUGtrue

\newif\ifBLINDED
\BLINDEDfalse

\newif\ifARXIV
\ARXIVfalse

\input{misc/typesetting}

\usepackage{acmart-taps}
\usepackage{xcolor}
\usepackage{color}
\usepackage{rotating}
\usepackage{tabularray}
\usepackage{pbox}
\usepackage{svg}

\usepackage{scalerel}
\usepackage{soul}
\usepackage[frozencache,cachedir=./pyg/]{minted}
\usepackage{xcolor}
\usepackage{textcomp}
\usepackage{pmboxdraw}

\input{misc/listing_style}
\input{misc/data}

\begin{document}



\title{An Evaluation of the Use of Bounded Model Checking for Memory Safety Verification of Embedded Network Stacks}
\title{Evaluating the Use of Unit Proofing for Memory Safety Verification of Embedded Software}
\title{Evaluating the Use of Unit Proofs for Verifying Memory Safety of Embedded Software}
\title{An Empirical Measurement Study of Compositional Bounded Model Checking for Memory Safety Verification}
\title{Do Unit Proofs Work? An Empirical Study of Compositional Bounded Model Checking for Memory Safety Verification}

\author{Paschal C. Amusuo}
\email{pamusuo@purdue.edu}
\orcid{0000-0003-1001-525X}
\affiliation{%
  \institution{Purdue University}
  \city{West Lafayette}
  \state{Indiana}
  \country{USA}
}

\author{Owen Cochell}
\email{cochello@msu.edu}
\orcid{0009-0003-5027-7710}
\affiliation{%
  \institution{Michigan State University}
  \city{East Lansing}
  \state{Michigan}
  \country{USA}
}

\author{Taylor Le Lievre}
\email{tlelievr@purdue.edu}
\orcid{0009-0007-4816-9875}
\affiliation{%
  \institution{Purdue University}
  \city{West Lafayette}
  \state{Indiana}
  \country{USA}
}

\author{Parth V. Patil}
\email{patil185@purdue.edu}
\orcid{0009-0005-7337-1114}
\affiliation{%
  \institution{Purdue University}
  \city{West Lafayette}
  \state{Indiana}
  \country{USA}
}

\author{Aravind Machiry}
\email{amachiry@purdue.edu}
\orcid{0000-0001-5124-6818}
\affiliation{%
  \institution{Purdue University}
  \city{West Lafayette}
  \state{Indiana}
  \country{USA}
}

\author{James C. Davis}
\email{davisjam@purdue.edu}
\orcid{0000-0003-2495-686X}
\affiliation{%
  \institution{Purdue University}
  \city{West Lafayette}
  \state{Indiana}
  \country{USA}
}

\begin{abstract}

Memory safety defects pose a major threat to software reliability, enabling cyberattacks, outages, and crashes.
To mitigate these risks, organizations adopt Compositional Bounded Model Checking (BMC), using unit proofs to formally verify memory safety. 
However, methods for creating unit proofs vary across organizations and are inconsistent within the same project, leading to errors and missed defects. 
In addition, unit proofing remains understudied, with no systematic development methods or empirical evaluations.

This work presents the first empirical study on unit proofing for memory safety verification.
We introduce a systematic method for creating unit proofs that leverages verification feedback and objective criteria. 
Using this approach, we develop \numUnitProofs unit proofs for four embedded operating systems and evaluate their effectiveness, characteristics, cost, and generalizability. 
Our results show unit proofs are cost-effective, detecting 74\% of recreated defects, with an additional 9\% found with increased BMC bounds, and 19 new defects exposed. 
We also found that embedded software requires small unit proofs, which can be developed in 87 minutes and executed in 61 seconds on average. 
These findings provide practical guidance for engineers and empirical data to inform tooling design.

\end{abstract}

\keywords{Empirical Software Engineering; Memory Safety; Software Verification; Bounded Model Checking; Cybersecurity; Embedded Software}


\setcopyright{none} 
\settopmatter{printacmref=false} 
\renewcommand\footnotetextcopyrightpermission[1]{}

\maketitle

\section{Introduction}
Memory safety defects~\cite{aleph_one_smashing_nodate} remain a significant threat to the reliability of critical software systems, particularly in programs written in C and C++. 
These defects facilitate cybersecurity attacks~\cite{noauthor_project_nodate}, software crashes~\cite{noauthor_memory_nodate} and outages~\cite{noauthor_widespread_2024}. 
A notable example is the July 2024 CrowdStrike outage~\cite{noauthor_widespread_2024}, caused by an out-of-bounds read~\cite{noauthor_channel_nodate}, affecting 8.5 million systems~\cite{weston_helping_2024} and resulting in a \$10 billion loss~\cite{wee_here_nodate}.
Additionally, these defects account for up to 70\% of high-severity vulnerabilities~\cite{alex_rebert_safer_nodate, msrc_team_proactive_nodate} in C/C++ codebases.
While memory-safe languages like Rust exist, many active C/C++ projects remain, with barriers to transition, and many Rust projects still depend on C/C++ libraries and unsafe code~\cite{sharma_rust_2024, fulton_benefits_2021}.
Given the severity of these issues, government agencies~\cite{bob_lord_urgent_2023, noauthor_darpa_nodate}, academia~\cite{watson_it_2025}, and industry~\cite{alex_rebert_securing_nodate} increasingly advocate for memory-safe development practices and stricter safety standards. 
Particularly, the United States Department of Defense encourages the use of formal methods to mitigate these critical defects~\cite{noauthor_darpa_nodate}.

\begin{figure}[h]
    \centering
    \includegraphics[width=0.7\linewidth, trim={1.5cm 2cm 1.5cm 1.5cm}, clip]{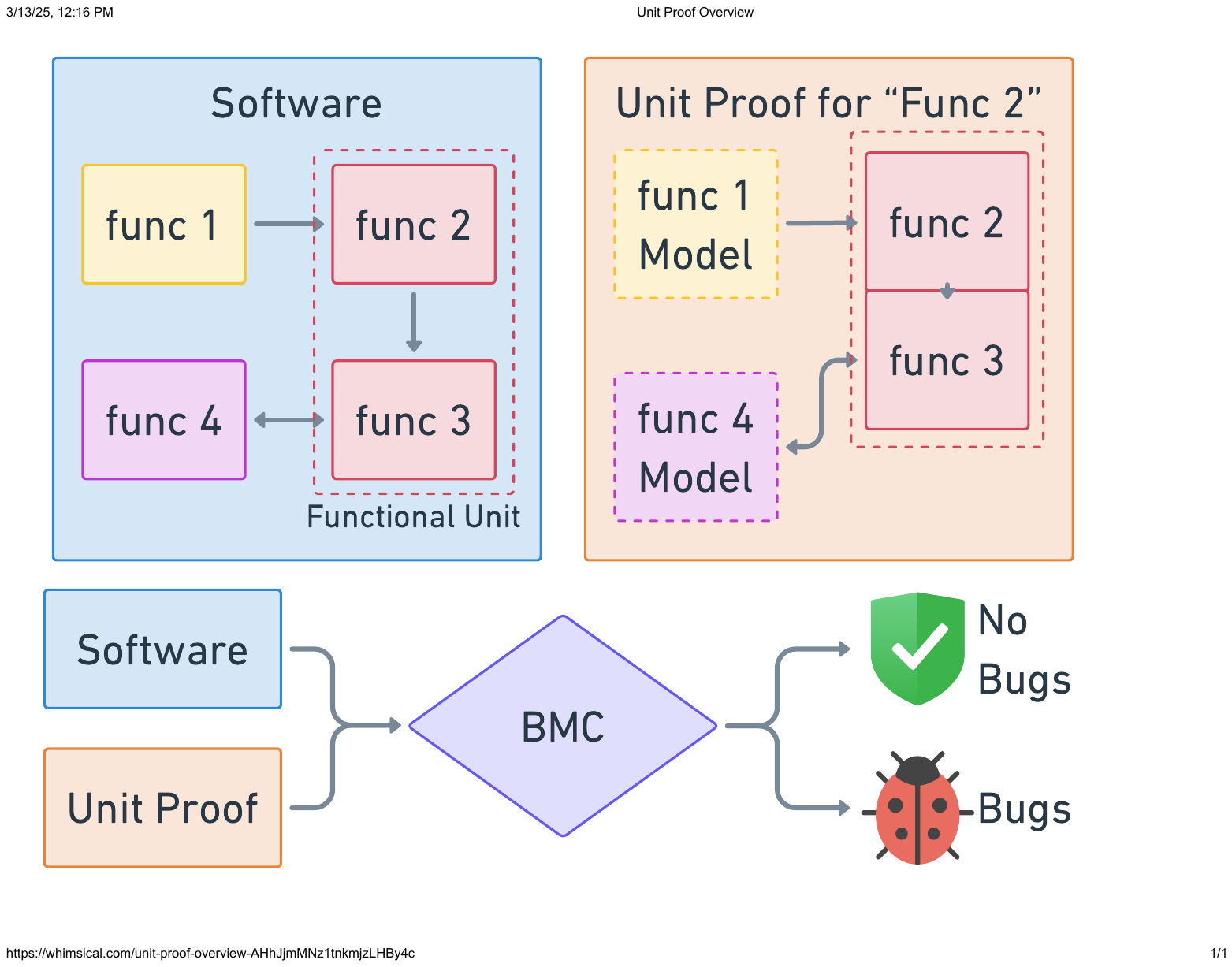}
    \caption{A unit proof, containing models for func1 and func4, verifies the functional unit comprising func2 and func3.}

    \label{fig:unit-proofing-overview}
\end{figure}

To mitigate memory safety defects, AWS~\cite{chong_code-level_2020}, ARM~\cite{wu_verifying_2024}, and Ant Group~\cite{wang_unsafecop_2025} employ Bounded Model Checking (BMC) to formally verify memory safety of software systems.
They adopt a compositional approach, known as unit proofing~\cite{amusuo_enabling_2024}, to verify functional units individually. 
This approach relies on \textit{unit proofs} that correctly model a unit's environment for verification. 
However, there is no standardized method for creating unit proofs to ensure correctness. 
Different organizations employ varying methodologies, leading to inconsistencies even within the same projects. 
The lack of a systematic approach introduces errors in unit proofs (\cref{subsec:rq4-results}), prevents the detection of defects~\cite{amusuo_enabling_2024}, and increases the cost of adoption for new organizations. 
In addition, unit proofing remains understudied in research.
Existing experience reports~\cite{chong_code-level_2020, wu_verifying_2024, wang_unsafecop_2025} only provide high-level descriptions without empirical validation. 
Prior studies on formal verification~\cite{matichuk_empirical_2015, olszewska_specification_2010, beyer_software_2017} do not address unit proofing or memory safety verification. 
As a result, no systematic methods exist for developing unit proofs and there are no empirical evaluations of unit proofing's effectiveness or costs.

This work presents the \ul{first} empirical study on unit proofing for memory safety verification and addresses five research questions. 
We evaluate on embedded software due to its safety-critical applications~\cite{feiler_challenges_2009} and the challenges of adopting alternative techniques (\cref{subsubsec:method-software-selection}).
First, we introduce a systematic method for unit proof creation, where coverage and error reports guide the unit proof development, and objective criteria are used to assess completion.
Using this method, we created \numUnitProofs unit proofs to verify functional units in four embedded operating systems. 
Using a blind approach (\cref{subsubsec:method-rq1}), we then reintroduce \recDefects known memory safety defects in these units and evaluate the effectiveness of unit proofs in detecting the recreated defects, analyze the unit proof characteristics and measure the cost of developing and using them.
Finally, we compare them with expert-developed unit proofs and assess if the systematic approach will generalize to unverified functions.

Our results provide empirical measurements of the benefits and costs of using unit proofs for memory safety verification. 
Of \recDefects recreated defects, systematic unit proofs detected 66 (74\%) and an additional 8 (9\%) with increased BMC bounds, while 10 remained undetected due to memory exhaustion. 
Additionally, unit proofs exposed 19 new defects, highlighting their practical impact. 
On cost, developing each unit proof took an average of \developmentTime minutes, with execution averaging \executionTime seconds. 
The median unit proof was developed in 72 minutes, executed in 25 seconds, and verified 185 lines of C code. 
Our results also show that 
    unit proof sizes correlate with functional unit size ($R^2 = \funcUnitrsq$);
    that
    verification formula size better predicts execution time than does program size ($R^2 = \formulaSizersq$ \textit{vs.} $R^2 = \programSizersq$);
    and that
    systematic unit proofs are smaller, faster, and achieve better coverage than those developed by project experts. 

In summary, our main contributions are:
\begin{itemize}
    \item A systematic approach to developing unit proofs for memory safety, including strategies for deriving unit proof models and loop bounds.
    \item The first empirical study on the use of unit proofs for memory safety verification. Our results provide evidence that unit proofs are cost-effective for ensuring memory safety.
    \item A dataset containing \numUnitProofs unit proofs, \recDefects recreated and \newDefects new memory safety defects to facilitate research on unit proofing.
\end{itemize}

\noindent
\textbf{\ul{Significance}}:
As software has become more critical to society, formal methods are now seeing broader industry adoption.
Our investigation aims for two benefits: 
    (1) to help engineering teams to develop effective unit proofs, and 
    (2) to guide research on automating unit proof development, thereby reducing unit proofing costs and enabling broader adoption of memory safety verification in software engineering practices.

\section{Background \& Related Work}
\label{sec:background}

In this section, we cover the memory safety concept and how it can be verified using Bounded Model Checking (BMC). 
Then we discuss approaches for applying BMC to larger software and review the empirical studies on formal verification.

\subsection{Memory Safety}
\label{subsec:bg-memory-safety}
The C and C++ languages allow software engineers to manipulate memory directly without restrictions. 
While this provides flexibility and performance benefits, it requires engineers to ensure \textit{memory safety} --- that no memory is accessed unless it has been explicitly allocated and with valid scope~\cite{szekeres_sok_2013}. 
Violating this property poses significant risks, including denial of service, information leaks, and remote code execution~\cite{szekeres_sok_2013, aleph_one_smashing_nodate}. 
Such defects are the leading causes of security vulnerabilities in memory-unsafe codebases~\cite{alex_rebert_safer_nodate, msrc_team_proactive_nodate} and are frequently exploited in zero-day attacks~\cite{noauthor_project_nodate}.

Szekeres~\etal~\cite{szekeres_sok_2013} describe various ways memory safety defects arise in practice. 
First, a memory reference is manipulated to point to an invalid memory location. 
This can result from untrusted input being directly used to compute the reference or from programming errors such as unchecked allocation failures and arithmetic errors that render the reference invalid. 
In the second step, this invalid reference is used to access memory, either for reading or writing. 
These steps may occur within the same function (\textit{intra-function}, \eg, in func2 of \cref{fig:unit-proofing-overview}), across different functions in the same functional unit (\textit{intra-unit}, \eg, in func2 and func3), or across distinct functional units (\textit{inter-unit}, \eg, in func2 and func4).


Several techniques are commonly used to detect or mitigate these defects, including 
    static analysis~\cite{shen_empirical_2023, pistoia_survey_2007}, 
    dynamic analysis~\cite{amusuo_systematically_2023, klooster_continuous_2023}, and
    runtime mitigation~\cite{davi_ropdefender_2011, cowan_protecting_nodate, clements_aces_2018, ganz_aslr_2017}.
In this work, we provide empirical measurements of a complementary and emerging formal technique for mitigating memory safety defects.

\begin{figure}[h]
    \centering
    \begin{minipage}{0.48\linewidth}
        \begin{minted}[
            fontsize=\scriptsize,
            linenos,
            gobble=2,
            xleftmargin=0.5cm,
            escapeinside=||,
            style=colorful,
            breaklines
        ]{c}
// Source Program
struct context {
    uint8_t payload[CONSTANT];
};

int targetFunc(char *data, size_t len) {
    context *ctx = get_current_ctx();
    for (i=0; i<3; i++) {
        ...
    }
    memcpy(ctx->payload, data, len);
    |\textcolor{red}{\textit{\_assert(sizeof(ctx->payload) >= len;)}}|

}
        \end{minted}
        \captionsetup{justification=centering}
        \label{code:memory-unsafe-program}
    \end{minipage}
    \hfill
    \begin{minipage}{0.48\linewidth}
        \begin{minted}[
            fontsize=\scriptsize,
            linenos,
            gobble=2,
            xleftmargin=0.5cm,
            escapeinside=||,
            style=colorful,
            breaklines
        ]{c}
// Unit Proof
context *get_current_ctx() {
    context *ctx = malloc(sizeof(context));
    |\textcolor{blue}{\textit{\_assume(ctx != NULL);}}|
    return ctx;
}

|\textcolor{purple}{\textit{\_unwind(targetFunc.0:2);}}|

void harness() {
    size_t len;
    |\textcolor{blue}{\textit{\_assume(len <= CONSTANT);}}|
    char *data = malloc(len);
    |\textcolor{blue}{\textit{\_assume(data != NULL);}}|
    targetFunc(data, len);
}
        \end{minted}
        \captionsetup{justification=centering}
        \label{code:unit-proof}
    \end{minipage}
    \label{code:unsafe-program-unit-proof}
    \captionof{listing}{The left program contains a potential memory safety defect and an expected memory safety property (in red). The right shows a unit proof used to verify the program. It contains preconditions that model input variables (in blue) and loop unrolling bounds (in purple).}
\end{figure}


    
    



\subsection{Bounded Model Checking for Verifying Memory Safety}
\label{subsec:bg-bmc}

Bounded Model Checking (BMC) is a verification technique that mathematically checks whether a program, within specified \textit{bounds}, satisfies a given property~\cite{clarke_bounded_2001}. 
Properties can be \textit{specific} to the program or \textit{agnostic} (\eg memory safety properties). 
A program can be bounded by restricting the number of loop iterations, recursion depth, size of data structures or its access to children functions.
To verify a property, BMC encodes the program and the specified property as a Boolean Satisfiability (SAT) problem and solves it using SAT/SMT solvers to identify any sequence of instructions within the program that can violate the specified property~\cite{martin_brain_cbmc_nodate}. 

Various BMC tools, such as the C Bounded Model Checker (CBMC)\cite{kroening_cbmc_2014} and Kani\cite{kani_getting_nodate}, automate memory safety verification by inserting assertions at memory access points (see red line in \cref{code:unsafe-program-unit-proof}) to verify the safety of these operations.
These assertions are translated into properties and analyzed by constraint solvers.
Additional tooling is also available to support engineers using BMC tools.
For example, the CBMC viewer~\cite{noauthor_reference_nodate} converts CBMC verification and coverage reports into browsable HTML and JSON formats.
The CBMC Proof Debugger~\cite{noauthor_model-checkingcbmc-proof-debugger_2024}, a VS Code plugin, aids in debugging error traces.
The Goto Harness~\cite{noauthor_cprover_nodate} assists in generating initial unit proofs for specific functions.
Our work provides an empirical study of compositional bounded model checking, an emerging technique for verifying memory safety.

\subsection{Scaling BMC to Software of Realistic Sizes}
\label{subsec:scaling-bmc}


\subsubsection{Compositional BMC}
While BMC is effective for small programs, its scalability is limited by the NP-completeness of Boolean satisfiability problems~\cite{schaefer_complexity_1978}. 
Compositional (or modular) BMC~\cite{sun_compositional_2008} addresses this limitation by decomposing programs into smaller, verifiable units, verifying each unit individually, and deriving guarantees for the entire program compositionally. 
This can be achieved through two approaches~\cite{kleine_buning_refined_2022}.
Specification-based methods use predefined specifications to guide a unit's verification, allowing each unit to be verified independently.
Conversely, fully automated approaches~\cite{cho_blitz_2013, beckert_modular_2020, kleine_buning_automatic_2019, hashimoto_modular_2009} decompose, verify, and integrate verification results without human intervention.
While compositional BMC enables scaling BMC to larger software, it can potentially miss \textit{inter-unit defects} where the memory reference invalidation and invalid memory access occur in different units (\cref{subsec:bg-memory-safety}).

\subsubsection{Unit proofing}
\label{subsubsec:bg-unit-proofing}
Unit proofing is a specification-based approach for compositional BMC where \textit{unit proofs} (the specification) are developed and used to verify individual software units.
Amusuo~\etal~\cite{amusuo_enabling_2024} describe this engineering activity of developing, using and maintaining unit proofs as ``unit proofing''.
Many organizations~\cite{chong_code-level_2020, wu_verifying_2024, wang_unsafecop_2025} have reported using this approach to verify the memory safety of critical software components.
As shown in \cref{fig:unit-proofing-overview}, unit proofs serve as function-level harnesses, modeling the unit’s environment (input variables, shared state, and undefined functions) and specifying necessary BMC bounds (loop bounds, unit scope~\footnote{A unit's scope is the source files compiled with the unit proof. Function definitions in these files are included during verification. Other functions remain undefined.}, data structure bound~\footnote{bound on array and linked-list sizes.}). 
\cref{code:unsafe-program-unit-proof} presents a sample unit proof for the \texttt{targetFunc} program, where blue lines model input variables, and the \texttt{get\_current\_ctx()} function models an actual implementation. 
Unit proofing is conceptually similar to unit testing, as both validate isolated software units. 
However, unlike unit tests that check correctness for specific inputs, unit proofs verify correctness for all inputs satisfying the provided models.

While multiple organizations have reported using unit proofs, their approach to creating the unit proofs vary: AWS~\cite{chong_code-level_2020} develops unit proofs from scratch, while ARM~\cite{wu_verifying_2024} and Ant Group~\cite{wang_unsafecop_2025} derive them from existing function specifications and unit tests, respectively. 
In all cases, unit proof creation was done in collaboration with development teams, with AWS iteratively refining models until they were accepted as accurate. 
Additionally, AWS and Ant Group reported increased bug detection rates due to unit proof adoption. 
However, none of these organizations disclosed detailed information on the models and BMC bounds used, how they derived them or the cost of developing the unit proofs.

\subsection{Empirical Studies on BMC}
\label{subsec:bg-empirical-evals}



Few studies have evaluated the effectiveness and cost of Bounded Model Checking (BMC). 
Unit proofing experience papers~\cite{chong_code-level_2020, wu_verifying_2024, wang_unsafecop_2025} report discovering new defects but lack a systematic validation of effectiveness. 
Beyer~\etal~\cite{beyer_software_2017} showed BMC outperforms fuzzing and testing in detecting defect on software benchmarks, but their study did not account for challenges in realistic software, where loop bounds~\cite{huang_loop_2025} or incorrect models~\cite{amusuo_enabling_2024} can hinder detection. 
Thus, an empirical evaluation of BMC in real-world software or through unit proofing remains missing.  

Regarding cost, few studies assess the cost of applying BMC to realistic software. 
AWS reported that a new engineer required one month to develop high-quality unit proofs and verify 1,500 lines of code~\cite{chong_code-level_2020}. 
Additionally, Huang~\etal~\cite{huang_lessons_2024} found that most formal verification projects required over a year of effort and verification expertise. 
Given the huge cost of formal verification, empirically measuring unit proofing costs is therefore essential for assessing its practical adoption.


\section{Research Questions}

We summarize as two knowledge gaps

\myparagraph{Gap 1: Lack of a standardized unit proofing approach} 
Reports~\cite{chong_code-level_2020, wu_verifying_2024, wang_unsafecop_2025} show organizations develop unit proofs using different methods, with inconsistencies even within the same project (\cref{subsec:rq4-results}). 
This increases error likelihood and compromises formal verification guarantees~\cite{amusuo_enabling_2024}.
Additionally, the absence of a standardized approach raises adoption costs, as new organizations may struggle to identify effective or cost-efficient methodologies.

\myparagraph{Gap 2: No empirical evaluations of unit proofing}
There has been no empirical assessment of the effectiveness, cost, or limitations of unit proofing for memory safety verification. 
Existing studies on formal verification~\cite{matichuk_empirical_2015, tsai_comparative_2000, olszewska_specification_2010,  beyer_software_2017} either do not cover memory safety verification or focus on benchmark programs which will not account for defects that span multiple functional units. 
Consequently, engineering teams lack data on the benefit-cost tradeoffs of unit proofing, hindering informed adoption decisions.


Given the variations in unit proofing approaches, their sometimes costly need for project expertise, and the absence of benefit-cost measurements, this work investigates the following question:
\textit{If we develop unit proofs using a uniform and systematic approach, would they expose memory safety defects? If so, what do they cost?}
We divide this into five \textbf{research questions} across three themes:

\vspace{0.05cm}
\noindent
\textbf{Theme 1: Unit proof effectiveness}

\begin{itemize} [leftmargin=30pt, rightmargin=5pt]
    
    \item [\textbf{RQ1:}] Does a systematic unit proofing approach enable detection of memory safety defects?

\end{itemize}

\noindent
\textbf{Theme 2: Unit proof characteristics and cost}

\begin{itemize} [leftmargin=30pt, rightmargin=5pt]
    
    \item [\textbf{RQ2:}] What are the characteristics of the unit proofs produced?
    
    \item [\textbf{RQ3:}] How long does it take to develop and execute unit proofs?  

    \item [\textbf{RQ4:}] How do unit proofs developed systematically differ from those developed by project experts?

\end{itemize}

\noindent
\textbf{Theme 3: Generalizability of approach}

\begin{itemize} [leftmargin=30pt, rightmargin=5pt]
    
    \item [\textbf{RQ5:}] Will the systematic unit proofing approach generalize across other functions? 
    
\end{itemize}

This paper presents a methodology that uses verification feedback to guide unit proof development and a set of objective criteria to assess completeness and quality.
Theme 1 evaluates whether unit proofs developed with this approach expose memory safety defects. 
Theme 2 examines their characteristics and the cost of developing and using the unit proofs. 
Theme 3 investigates the generalizability of this approach beyond the studied sample.

\section{Methodology}


\subsection{Study Design}

\begin{figure*}[h]
    \centering

    \includegraphics[width=\linewidth, trim={1.2cm 11cm 1.3cm 3cm}, clip]{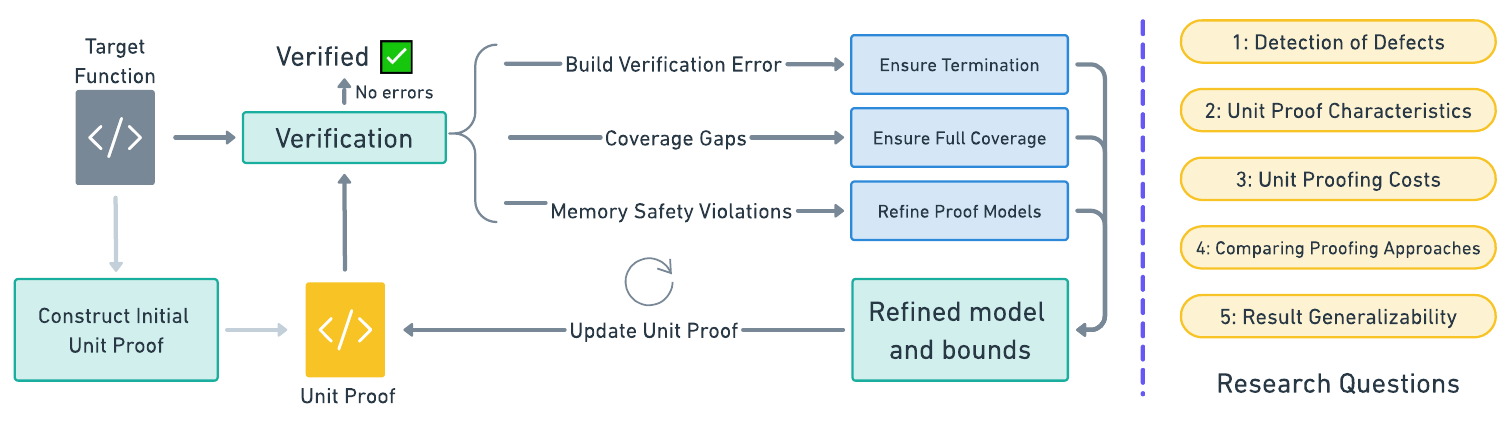}
    \caption{
    Study Methodology. We first develop unit proofs in 5 steps, iterating until full coverage and no memory safety violations. Then we evaluate the five research questions.
    }

    \label{fig:unit-proofing-method}
\end{figure*}

\subsubsection{Software selection}
\label{subsubsec:method-software-selection}



We conducted our study on four open-source embedded operating systems: FreeRTOS~\cite{freertos_freertos_nodate}, Zephyr~\cite{noauthor_zephyr_nodate}, Contiki-ng~\cite{oikonomou_contiki-ng_2022} and RIOT~\cite{noauthor_riot_nodate}. 
Embedded software is ideal for memory safety verification due to their use in safety-critical systems~\cite{feiler_challenges_2009}, lack of hardware mitigations~\cite{abbasi_challenges_2019}, fuzzing challenges~\cite{muench_what_2018, yun_fuzzing_2023} and high cost of fixing defects after deployment. 
The selected OSes are actively maintained and include diverse components (kernels, drivers, protocols, and libraries) that process external data, making memory safety defects in them exploitable.
We selected them among other OSes listed on www.osrtos.com~\cite{noauthor_osrtos_nodate} based on popularity (GitHub stars) and publicly disclosed vulnerabilities.

\subsubsection{Functional unit selection}
\label{subsubsec:method-unit-selection}
We selected functional units in the selected software with previously known defects.
First, we identified defects in the National Vulnerability Database (NVD) that affect the selected software, has a memory safety CWE (\eg CWE-125, CWE-476, CWE-787)\footnote{The NVD tracks security defects with a description, CVE identifier, and CWE category.}, and contained sufficient details for reproduction.
This yielded 152 defects (Zephyr - 69, Contiki-ng - 47, RIOT-OS - 28 and FreeRTOS - 8).
To reduce bias due to the varying number of defects, we selected 30 Zephyr defects and 25 Contiki-ng defects, as well as all RIOT and FreeRTOS defects, yielding a total of \recDefects defects.

To locate the affected functional unit for each defect, we reviewed the associated security advisories and patches to identify the lines of code where the defect manifests (an invalid memory is accessed) and was fixed.
We identify the affected functional unit's entry as the closest function that reaches both locations.
We selected \numUnitProofs unique functional units, found across \numEmbeddedComponents distinct embedded software components (\eg network subsystem, file system, USB driver, etc) and \numProtocols protocol implementations (\eg TCP, IPv6, DHCP, etc). 

\subsubsection{Bounded model checker selection:}
We conducted all bounded model checking using the Ansi-C Bounded Model Checker (CBMC v6.3.1)~\cite{kroening_cbmc_2014}. 
CBMC is actively maintained and used in industry~\cite{chong_code-level_2020}.

\subsubsection{Experimental setup:}
We used a Dell Precision 5820 Tower server: Intel(R) Xeon(R) W-2295 CPU @ 3.00GHz, 188GB of RAM. 

\subsection{Systematically Developing Unit Proofs}
\label{subsec:method-developing-unit-proofs}

We describe our approach, completeness criteria, and unit proofing steps. \cref{code:realistic-unit-proof} illustrates the steps.


\subsubsection{Unit proofing approach}
\label{subsubsec:method-approach}
The key difficulty in unit proofing is modeling the unknown components (variables, functions) a unit depends on (\cref{subsubsec:bg-unit-proofing}). 
We use a \textit{bottom-up strategy} called ``angelic modeling''~\cite{das_angelic_2015} and ``assume-guarantee reasoning''~\cite{cobleigh_breaking_2008}, which derives models that ensure the target unit behaves correctly, verifies the unit using these models, and then validates the model assumptions.
In this work, we systematically use the coverage report and error traces to model unknown components.
A similar approach has been applied in prior works~\cite{cobleigh_learning_2003, calcagno_compositional_2009, das_angelic_2015}.

Alternatively, a \textit{top-down strategy} first constructs accurate models of unknown components for verification, as seen in prior unit proofing reports~\cite{chong_code-level_2020, wu_verifying_2024, wang_unsafecop_2025}. 
However, it depends on existing specifications or project-wide knowledge, increasing its potential cost.

\subsubsection{Completeness criteria for unit proofs}
\label{subsubsec:method-assessing-unit-proofs}

We used the following objective criteria to assess the completeness of unit proofs.
As prior works do not provide any metrics for assessing unit proofs, we derived these metrics from our experience.

\begin{enumerate}
    \item \textit{Graceful Termination:} The unit proof builds, executes and terminates after verification completes. 
    \item \textit{Coverage:} The unit proof covers all reachable lines of code in the functional unit. 
    Line coverage is commonly used for assessing unit tests~\cite{zhu_software_1997, horgan_achieving_1994} and fuzzing campaigns~\cite{wang_be_2019, bohme_reliability_2022}.
    \item \textit{Memory safety violations:} 
    Our unit proofing philosophy relies on using memory safety violations in the error report to derive models of unknowns in the unit proof.
    Hence, we consider the unit proof complete when all reported memory safety violations are either traced to unknowns and resolved using refined models or they are identified as real bugs and fixed.
   
\end{enumerate}

\begin{figure}[h]
    \centering
    \includegraphics[width=\linewidth]{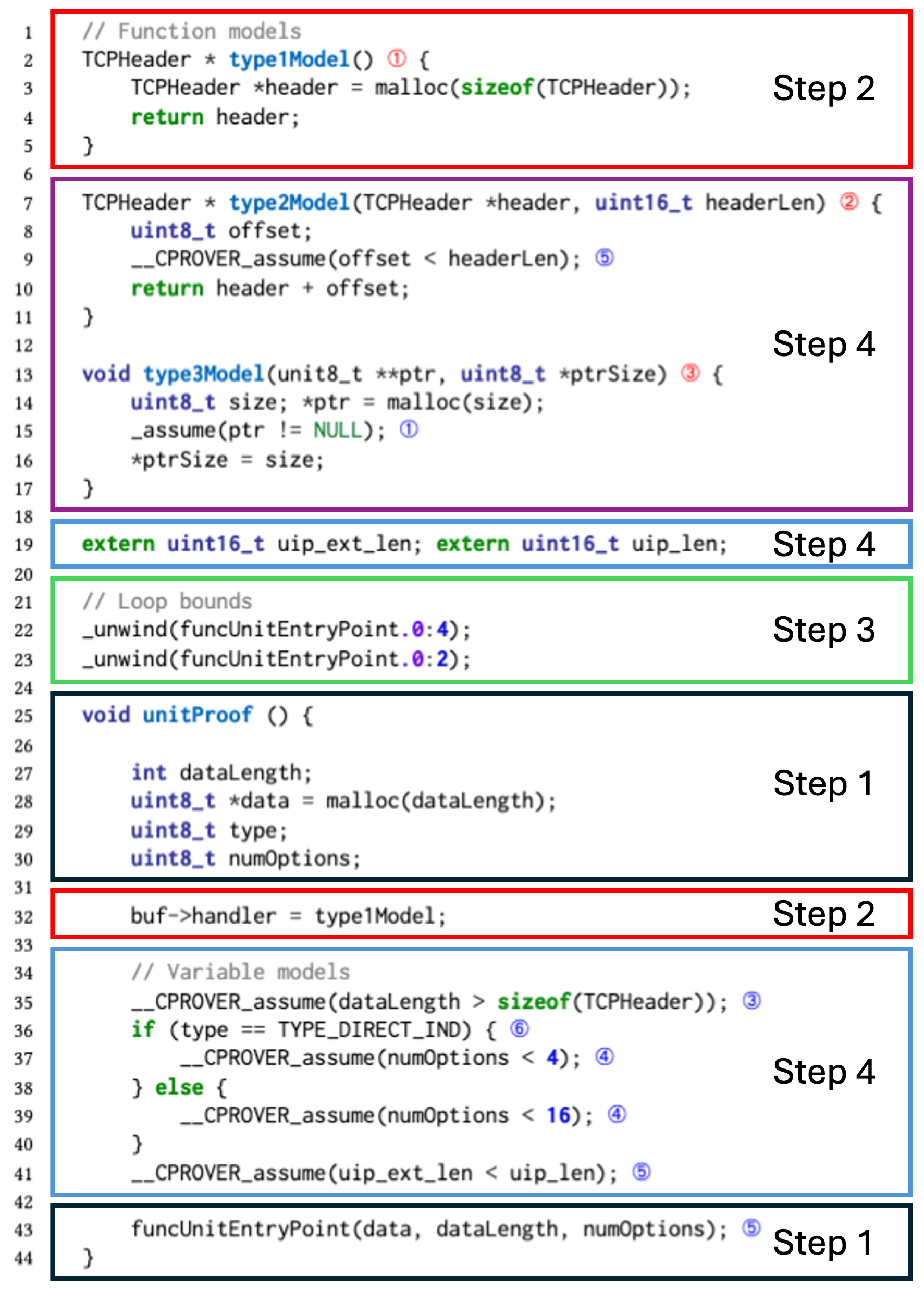}
    \caption{\small A systematically-developed unit proof. 
        For realism, each element is taken from a real proof.
        In step 1, the initial unit proof contained only the black box. 
        Initial verification timed out due to an unconstrained function pointer.
        In step 2, we resolved this (adding red box).
        The coverage report showed gaps due to insufficient loops.
        In step 3, we increased the bounds of affected loops (adding green box).
        Finally, the error report indicated errors caused by unknown variables and functions.
        In step 4, we resolved these via variable preconditions (blue box) and function models (purple box).
        }

    \label{code:realistic-unit-proof}
\end{figure}

\subsubsection{Step 1: Constructing an initial unit proof}

First, we set up the proof directory containing the proof harness and proof makefile. 
The initial proof harness is based on the CBMC's proof writing guide~\cite{noauthor_cbmc-starter-kittraining-materialproof-writingmd_nodate} and defines input variables, allocates necessary pointers, and invokes the target function (entry point of the functional unit). 
The makefile specifies build (\eg program configurations) and verification (\eg loop bounds) options. 
We reuse the Makefile from the FreeRTOS repository~\cite{noauthor_freertosfreertos_2025} and set all loop bounds to 1.

\subsubsection{Step 2: Ensuring termination}
\label{subsubsec:method-step-2-termination}

We build and execute the unit proof using the \texttt{make} command. 
However, the process may fail or lead to prolonged verification. 
Build failures typically result from missing variable definitions, header files, or incorrect configurations. 
We analyze error messages and resolve them accordingly. 
For prolonged verification or memory exhaustion, we identify the causes and apply appropriate interventions. 
Common interventions involved initializing function pointers, breaking recursion chains and replacing a complex called function with a simpler model.
Our artifact (\cref{sec:data-availability}) contains details and examples. 

\subsubsection{Step 3: Ensuring full coverage}

After verification, we identify uncovered code blocks from the coverage report.
These coverage gaps were typically caused by insufficient loop bounds or data allocation, and fixed by increased loop unwinding on affected loops or setting minimum bound on data allocation.

\subsubsection{Step 4: Refining variable and function models}

Memory safety violations, reported during verification, may be caused by unknown components and not represent actual defects.
In such cases and following our angelic modeling approach (\cref{subsubsec:method-approach}), we use the reported error traces to identify which unknown variable or function caused the error and derived or refined the model for that component.

For violations linked to input or global variables, we derive the minimal precondition necessary to resolve the issue. 
If violations stem from an undefined function's return value, we introduce a model for the function and constrain the return value.
In instances where the unknown variable is initialized or validated in an undefined function (these functions had descriptive names like \texttt{prvAllowIPv4Packet} in FreeRTOS), we model the function's expected behavior or expand the unit’s scope by including the file containing the missing function's definition. 
Examples of derived models are in \cref{code:realistic-unit-proof} and in our artifact (\cref{sec:data-availability}).

After deriving preconditions, we then analyze the functions producing these variables (\eg callers or undefined functions) to determine if the derived precondition can be violated.
Without automated tools, we prioritize unlikely preconditions based on experience and intuition.
If a violation was caused by an unvalidated local field (\eg field read from an input data) or the derived precondition can be violated, we report it as a defect to maintainers.

\subsection{Answering Research Questions}
\label{subsec:method-research-questions}

\subsubsection{RQ1: Detection of memory safety defects}
\label{subsubsec:method-rq1}


We evaluate whether unit proofs developed systematically, using the approach in \cref{subsec:method-developing-unit-proofs}, can detect memory safety defects.
We re-introduce known memory safety defects in their corresponding functional units, assess the defect detection rate and investigate reasons for non-detection.

\myparagraph{Re-introducing known defects}
Similar to prior work~\cite{do_hardfs_2013, amusuo_systematically_2023}, we reintroduce the defects selected in \cref{subsubsec:method-unit-selection} by identifying the patches that fixed them and reverting these changes in the affected embedded OS.
Identifying patches involved reviewing security advisories and relevant GitHub pull requests to understand how each defect manifests and was fixed. 
Since most defects were addressed by adding validation, reversion typically involved removing or falsifying these checks. 
Initially, we attempted reverting to pre-fix versions but encountered frequent build failures due to unavailable, incompatible, or outdated dependencies, making this approach impractical and difficult to scale.

\myparagraph{Assessing detection of defect}
To determine if a defect is exposed, we execute the corresponding unit proof and analyze the error report for related memory safety violations. 
If no violation is reported, we re-examine the affected function and execution logic, modifying the unit proof as needed to expose the defect. 
We report any required interventions and reasons for non-exposure.

\myparagraph{De-biasing}
Clearly if an analyst knows the details of a defect, their approach may be biased.
To mitigate this, we used a blind approach when creating the unit proofs. 
One author identified the functions and introduced the defects while all unit proofs were developed by other authors who received only the affected functions.
The remaining issue is that the proof-writers knew that the functions were indeed defective, which may have made them more persistent.
We note two factors that mitigate this:
  (1) our systematic approach gave them clear stopping criteria, causing them to miss some defects, and
  (2) our approach caused them to find new defects in \newDefects of the \numUnitProofs functions they verified, indicating that they kept an open mind.


\subsubsection{RQ2: Characteristics of unit proofs}
\label{subsubsec:method-rq2}

We characterize unit proofs developed systematically by their size and contents.
A program's size and contents impacts its understandability~\cite{lin_evaluation_2008, boehm-davis_chapter_1988} and likelihood to contain errors~\cite{antinyan_mythical_2018}.
Hence, smaller, simpler unit proofs will be easier for software engineers to develop, understand, and maintain.
Understanding the models and loop bounds required in unit proofs can also inform future research on automating their derivation.
We measure \textit{size} by counting lines of code and number of models in the unit proof. 
To analyze \textit{structure}, we categorize model types and loop bounds used in the unit proof. 

\subsubsection{RQ3: Time to develop and execute unit proofs}
\label{subsubsec:method-rq3}

We evaluated the time required to develop and execute unit proofs using the proposed methodology. 
The time to develop unit proofs will affect adoption costs and influence adoption decisions, while execution time will impact the feasibility of integrating unit proofs during continuous software development. 
Given our iterative approach, the execution time will also influence the development time and the overall verification experience.  

For \timedUnitProofs unit proofs~\footnote{These were the last unit proofs we developed. We used the first 52 proofs to develop the necessary unit proofing experience.}, we measure and report development time for each proof and each step of the process (\cref{subsec:method-developing-unit-proofs}) and identify the reasons for longer durations. 
We then measure verification time for all unit proofs by modifying the build file to report the duration of the \texttt{cbmc} execution command, and excluding compilation time. 
Finally, we examine how execution time correlates with program size and the complexity of the solved formula to assess their potential as predictors of verification time.



\subsubsection{RQ4: Systematic vs expertise-based unit proofs}

We evaluate how unit proofs developed systematically (\cref{subsec:method-developing-unit-proofs}) and with objective completeness criteria (\cref{subsubsec:method-assessing-unit-proofs}) compare to those developed by project experts.
Among the selected software, only FreeRTOS contained existing unit proofs.
We identified \numFreeRTOSProofs FreeRTOS unit proofs covering the same functional units selected in \cref{subsubsec:method-unit-selection}. These proofs, developed in 2020, likely followed the process described by Chong~\etal~\cite{chong_code-level_2020}.
We then compared the characteristics (\cref{subsubsec:method-rq2}) and execution time (\cref{subsubsec:method-rq3}) of the FreeRTOS unit proofs with the corresponding systematic unit proofs.
Finally, we assessed how the different development approaches influenced these results.

\subsubsection{RQ5: Generalizability to other functions}

RQ1 to RQ4 were evaluated on a subset of functions from the selected software, which may not represent the remaining functions, limiting generalizability. 
To assess this potential bias, we compare the evaluated subset to the remaining functions in each software. 
Given that verification effort correlates with program size and complexity~\cite{olszewska_specification_2010, matichuk_empirical_2015}, we measure these metrics using the Lizard tool~\cite{yin_terryyinlizard_2025}. 
We collect lines of code and complexity data for all functions and created two datasets: one for functions with unit proofs and another for all functions. 
We then compare their size and complexity metric distributions using statistical plots.

\section{Results}

In this section, we present our results by RQ (\cref{code:simple-unit-proof}). 

\subsection{RQ1: Detection of Memory Safety Defects}
\label{subsec:rq1-results}

\begin{table}
    \centering
    \caption{
    Table showing the proportion of known vulnerabilities that were exposed by unit proofing, and the number of new vulnerabilities found.
    }
    \begin{tabular}{lcc}
    \toprule
        \textbf{Defect Status} & \textbf{Count (\%)} \\
        \midrule
        \textit{Total Exposed Systematically} & 66 (74\%) \\
        \midrule
        \textit{Total Exposed with Interventions} & 8 (9\%) \\
        \ \ Increasing loop bounds & 4 \\
        \ \ Expanding unit's scope & 3 \\
        \ \ Increasing bounds on string length & 1 \\
        \midrule
        \textit{Total Not Exposed} & 15 (17\%) \\
        \ \ Memory Exhaustion (before adding defect) & 4 \\
        \ \ Memory Exhaustion (after adding defect) & 6 \\
        \ \ Substantially modified code & 1 \\
        \ \ Undefined API & 2 \\
        \ \ Requires timer event & 1 \\
        \ \ Unknown & 1 \\
        \midrule
        \textit{Total known defects} & 89 (100\%) \\
        \midrule
        \midrule
        \textit{New Defects Found} & 19 \\
        \ \ Out-of-bound write & 5 \\
        \ \ Out-of-bound read & 12 \\
        \ \ Null pointer dereferencing & 1 \\
        \ \ Arithmetic underflow & 1 \\
        \bottomrule
    \end{tabular}
    \label{tab:rq1-vuln-detection-status}
\end{table}

\Cref{tab:rq1-vuln-detection-status} presents the results of using systematically developed unit proofs to detect memory safety defects. 
Of the 89 defects analyzed, 66 (74\%) were exposed using systematic unit proofs, 8 (9\%) required increasing BMC bounds, and 15 (17\%) remained undetected. Also, 19 new defects were exposed using the systematic unit proofs.

\myparagraph{Investigating defects requiring additional interventions}
Among the five defects requiring increased bounds (\eg CVE-2024-32017), four involved copying strings into fixed-size buffers. 
Our approach (\cref{subsec:method-developing-unit-proofs}) adjusted BMC bounds only when it was necessary to increase coverage, and since the string-copying operations were already covered, further increases were not triggered. 
One defect (CVE-2023-6749) also required a string length bound of 32, exceeding our default of 20. 
The last defect (CVE-2020-14935) also involved accessing a fixed size buffer in a loop and was not also exposed because the function's coverage only required a lower loop bound.

Three other defects (\eg CVE-2024-4785) were missed due to our systematic approach limiting unit scope to functions within a single source file. Since the invalid memory access occurred in children functions in separate files, detecting them required expanding the unit's boundaries to include the relevant external function.

\myparagraph{Investigating defects not exposed}
Ten defects were undetected because the bounded model checker (BMC) ran out of memory before finding a solution. 
Four occurred during initial unit proof development, before the defect was reintroduced. 
In three cases, the issue involved memmove, which, when removed, resolved the problem. 
The fourth case involved an undefined function, whose removal also resolved the problem, though the exact reason remains unclear.
In the remaining six cases, detecting each defect required expanding the unit's boundary to include an additional file. However, this expansion led to memory exhaustion, preventing verification.

Two defects stemmed from unsafe sprintf usage, but since CBMC does not support sprintf, they were not exposed. Two others were missed due to significant changes in the target function, as the defect manifests during timer event processing, which CBMC does not support. The cause of the final undetected defect remains unknown.

\myparagraph{Exposing new defects}
Using systematically developed unit proofs, we uncovered 19 new defects: 9 in RIOT, 5 in Zephyr, and 5 in Contiki-ng. 
No defects were found in FreeRTOS, likely because the functions we verified had already been verified by the FreeRTOS team.  
All defects were reported to maintainers following their responsible disclosure policies. Five have been fixed and assigned CVE identifiers (with CVSS scores above 8.0 indicating high severity). The rest are under investigation. Details of these defects are included in our artifact.

\definition{
\textbf{RQ1 Finding:} Unit proofs developed systematically detected 74\% of recreated defects and 19 new defects. Increasing BMC bounds exposed additional 9\%. 11\% of defects not exposed due to memory exhaustion during verification. 
}


\subsection{RQ2: Unit Proof Characteristics}
\label{subsec:rq2-results}

RQ2 characterizes the size and components of systematically developed unit proofs.

\subsubsection{Characterizing unit proof sizes}

\begin{figure}[h]
    \centering
    \includegraphics[width=0.85\linewidth]{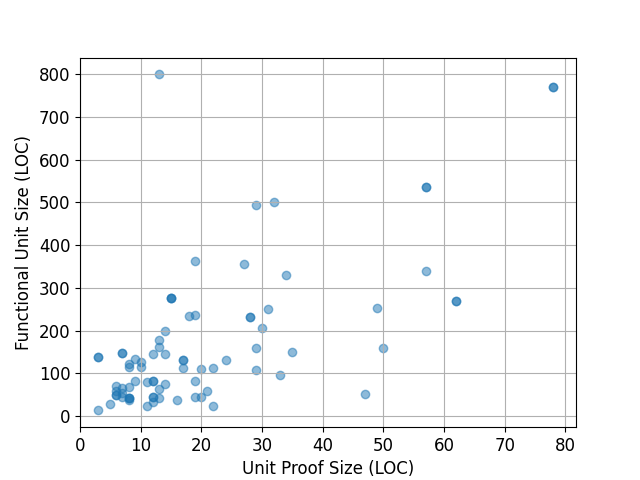}
    \caption{
    Plot showing unit proof sizes vary by unit sizes.
    }
    \label{fig:rq2-unit-proof-sizes}
\end{figure}

\begin{figure}[h]
    \centering
    \includegraphics[width=0.85\linewidth]{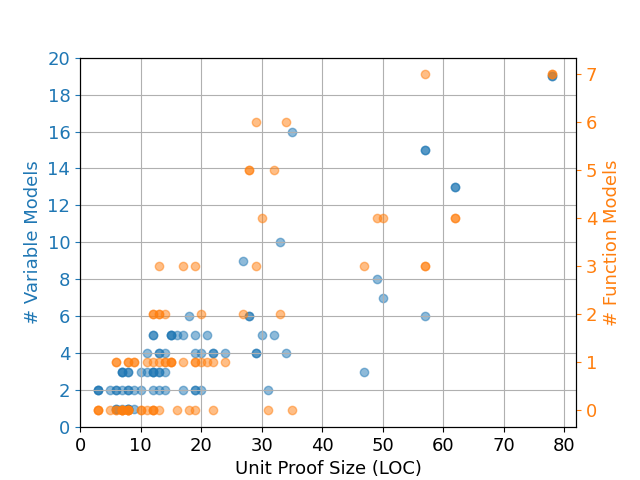}
    \caption{
    Plot showing the number of variables and models in unit proofs of different sizes.
    }
    \label{fig:model-unit-proof-size}
\end{figure}

\cref{fig:rq2-unit-proof-sizes} illustrates the size of unit proofs across different functional units. 
Notably, proofs had mean size of 20 lines of code, with only five exceeding 50 lines. 
The data indicates a direct correlation between unit proof size and functional unit size: larger units require proportionally larger proofs, averaging approximately one-tenth of the unit's size. 

\Cref{fig:model-unit-proof-size} shows the number of variable and function models in unit proofs of varying sizes. 
On average, unit proofs contain 5 variable and 2 function models, with model count increasing with proof size. 
These findings indicate that realistic functional units in embedded software typically require small unit proofs, typically less than 50 lines, and only a few models.

\subsubsection{Characterizing variable and function models}

\begin{table*}
    \centering
    \caption{\small
    Prevalence of different variable models across unit proofs. The different model types are numbered and illustrated in \cref{code:realistic-unit-proof}.
    }
    \begin{tabular}{lp{1.2cm}p{1.5cm}|p{1.5cm}p{1.7cm}p{1.5cm}p{1.5cm}p{1.3cm}p{1.8cm}}
    \toprule
        \textbf{OS} & \textbf{\# Unit Proofs} & \textbf{\# models} & \textbf{Pointer not null \dingone}& \textbf{Pointer and offset \dingtwo} & \textbf{Allocation size \dingthree} & \textbf{Integer range \dingfour} & \textbf{Integer r/ship \dingfive} & \textbf{Conditional modeling \dingsix} \\
        \midrule
        Zephyr & 25 & 159 & 87 & 8 & 31 & 15 & 16 & 2 \\
        Contiki-ng & 18 & 51 & 19 & 2 & 9 & 11 & 8 & 2 \\
        RIOT & 22 & 132 & 78 & 3 & 32 & 9 & 8 & 1 \\
        FreeRTOS & 8 & 25 & 17 & 0 & 6 & 2 & 0 & 0 \\
        \midrule
        Total & 73 & 367 & 201 (55\%) & 13 (4\%) & 106 (21\%) & 37 (10\%) & 32 (9\%) & 5 (1\%) \\
        \bottomrule
    \end{tabular}
    \label{tab:rq2-variable-models}
\end{table*}

\begin{table}
    \centering
    \caption{\small
    Characterizing function models in unit proofs. 
    Type 1 models model only the function's return value based on the type.
    Type 2 models model the return type using program-specific semantics.
    Type 3 models input arguments together with return values.
    These models are numbered and illustrated in \cref{code:realistic-unit-proof}.
    }
    \begin{tabular}{p{1.41cm}p{1.3cm}p{1.4cm}p{1.3cm}p{1.2cm}}
    \toprule
        \textbf{OS} & \textbf{\# models} & \textbf{Type 1 \dingrone} & \textbf{Type 2 \dingrtwo} & \textbf{Type 3 \dingrthree} \\
        \midrule
        Zephyr & 37 & 29 & 5 & 3 \\
        Contiki-ng & 21 & 14 & 6 & 1 \\
        RIOT & 40 & 34 & 4 & 2 \\
        FreeRTOS & 15 & 9 & 4 & 2 \\
        \midrule
        \textit{Total} & 113 & 86 (76\%) & 19 (17\%) & 8 (7\%) \\
        \bottomrule
    \end{tabular}
    \label{tab:rq1-function-models}
\end{table}

\cref{code:realistic-unit-proof} presents a realistic unit proof, highlighting the types of variable and function models it contains. 
\cref{tab:rq2-variable-models} shows that pointer-not-null \dingone and allocation size \dingthree models account for 76\% of variable models in unit proofs. 
As discussed in \cref{subsec:discussion-future-work}, these we observed that these models can be derived statically.
In other cases, recognizing these model types helped the authors determine how to model unknown variables to resolve reported violations.

\cref{tab:rq1-function-models} categorizes function models into three types: Type 1 and Type 2 model only the function’s return value, while Type 3 models include input behavior. Among 113 modeled functions, 76\% required only return value modeling, typically so it returns valid pointers of constrained sizes. This demonstrates that memory safety verification primarily relies on return value modeling, reducing the cost compared to verification tasks requiring full semantic behavior modeling.

\subsubsection{Characterizing loop bounds}
\label{subsubsec:result-rq2-characterizing-loops}

\begin{table}
    \centering
    \caption{
    Characterizing loops based on exit condition.
    }
    \begin{tabular}{lcccc}
    \toprule
        \textbf{Loop condition type} & \textbf{Count} & \textbf{b = 2} & \textbf{b = 3} & \textbf{max b} \\
        \midrule
        Constant & 33 & 9 & 12 & 65 \\
        Data length & 13 & 10 & 3 & 3 \\
        Linked-list length & 18 & 16 & 2 & 3 \\
        strlen/strcpy/memcmp & 21 & 0 & 0 & 65 \\
        Others & 7 & 5 & 1 & 5 \\
        \midrule
        Total & 92 & 40 (43\%) & 18 (20\%) & 65 \\
        \bottomrule
    \end{tabular}
    \label{tab:rq2-loop-types}
\end{table}

\cref{tab:rq2-loop-types} categorizes loops requiring custom bounds based on their exit conditions. 58 (63\%) loops required unrolling only 2 or 3 times. 
The required bound varied by loop type: loops with a static number of iterations required full unrolling, while loops processing variable-length data or linked lists typically needed a bound of 2 or 3. 
String-handling loops required bounds matching the maximum string length, and loops in \texttt{memcmp} implementations required bounds corresponding to the \texttt{memcmp} size argument.

\definition{
\textbf{RQ2 Finding:} Systematic unit proofs are small, averaging 20 lines of code, 5 modeled variables, and 2 modeled functions. 76\% of variable models can be derived statically, and 77\% of function models return unconstrained values. 63\% of loops require bounds <= 3, with the bound depending on loop type.
}

\subsection{RQ3: Unit Proof Development and Execution Time}
\label{subsec:rq3-results}

\begin{figure}[h]
    \centering
    \includegraphics[width=0.95\linewidth]{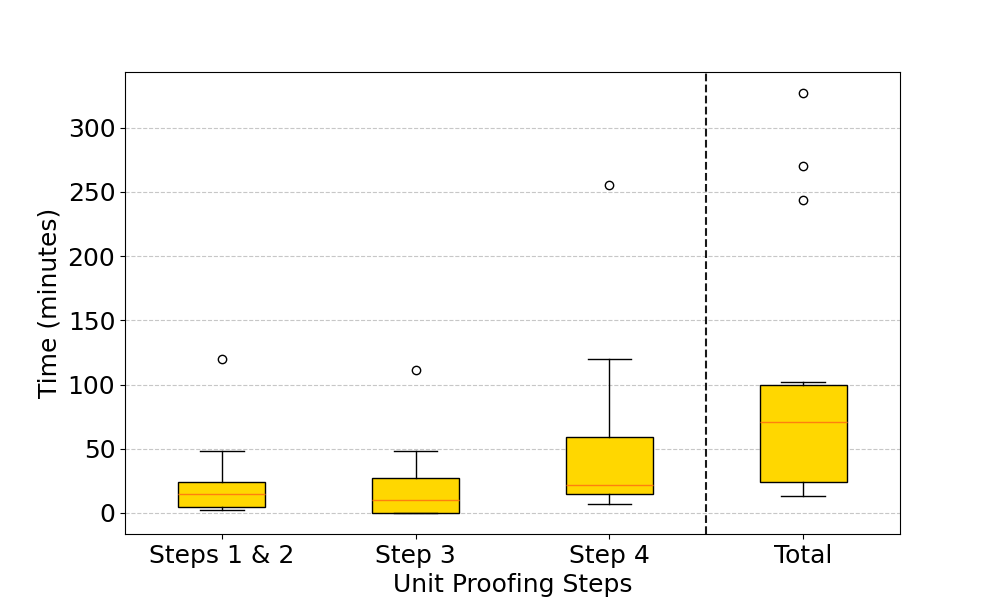}
    \caption{
    Time it takes to develop unit proofs.
    }

    \label{fig:rq3-verification-step-effort}
\end{figure}

\subsubsection{Unit Proof development time}

\cref{fig:rq3-verification-step-effort} illustrates the time spent on unit proof development, covering setup (steps 1 \& 2), coverage improvement (step 3), and model derivation (step 4). The median unit proof took 72 minutes, with significant time devoted to modeling unknown variables and functions based on reported violations.  

Delays in steps 1 \& 2 were mainly due to missing headers or undefined symbols when compiling units in isolation. 
Step 3 took longer for units with multiple loops, as our iterative approach derived loop bounds one at a time to improve coverage. 
Step 4 delays mostly arose from units with substantially larger sizes and which which required more modeled variables.
For instance, the two unit proofs with the longest Step 4 duration (4 hours 15 mins and 2 hours) verified 412 and 766 lines of code respectively and required 17 and 19 variable models respectively.

The reported durations exclude the one-time setup overhead for each embedded OS, which involved adapting to different build instructions, header paths, and configuration definitions. 
For example, Zephyr’s custom build system (West~\cite{noauthor_west_nodate}) relied on autogenerated headers and custom compile flags, causing errors when compiling files in isolation. 
We resolved this by modifying our Makefile to build a full Zephyr kernel, generating headers and extracting compile flags from compile\_commands.json. 
Once resolved, this approach was reused for other unit proofs within the same OS.



\begin{figure}[h]
    \centering
    \includegraphics[width=\linewidth]{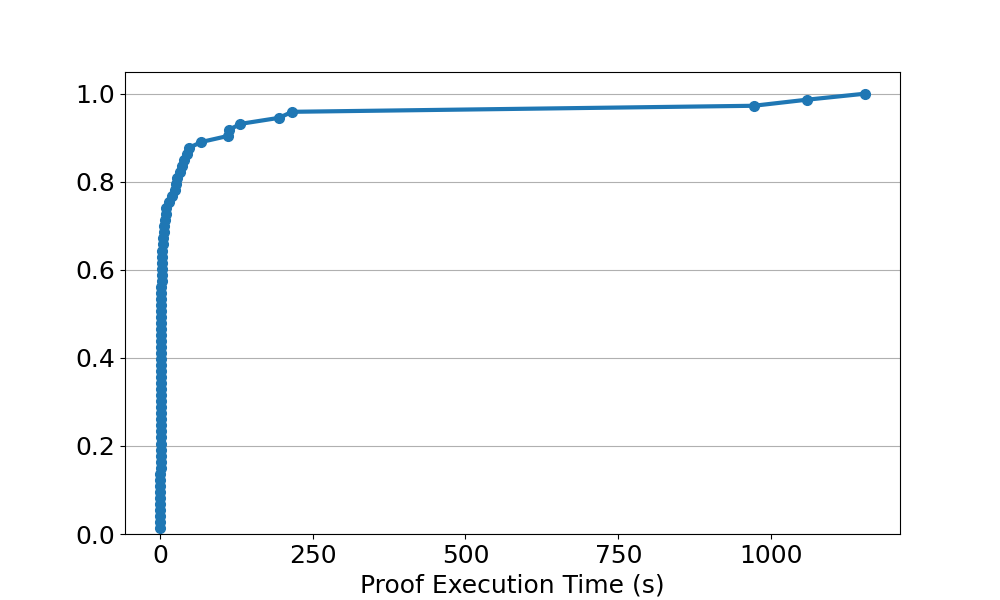}
    \caption{
    Cumulative density function (CDF) showing the time it takes to execute unit proofs.
    }
    \label{fig:rq3-verification-time}
\end{figure}

\subsubsection{Unit Proof execution time}

\Cref{fig:rq3-verification-time} shows unit proof execution times, with \proofsBelowSixtySecs\% completing in under 60 seconds and only 5 taking longer than 3 minutes. 
We attribute these fast times to minimal loop bounds, limiting the scope of units, and handling complex semantics like function pointers and recursion (\cref{subsubsec:method-step-2-termination}). 
This supports the feasibility of integrating unit proof execution into development or code commits for continuous verification~\cite{ohearn_continuous_2018}. 
When compared, we also found that execution time correlated more strongly with verification formula size ($R^2 = \formulaSizersq$) than program size ($R^2 = \programSizersq$) (Data and charts in artifact). 
Hence, they can be used to predict the verification time and help when determining appropriate size of a functional unit.

\definition{
\textbf{RQ3 Finding:} On average, unit proofs take 82 minutes to develop and 61 seconds to execute. The median proof required 72 minutes to develop, 25 seconds to execute, and verified 185 lines of code. Verification formula size accounts for 85\% of the execution time.
}

\subsection{RQ4: Systematic vs expertise-based Proofs}
\label{subsec:rq4-results}


\begin{figure}[h]
    \centering
    \includegraphics[width=\linewidth]{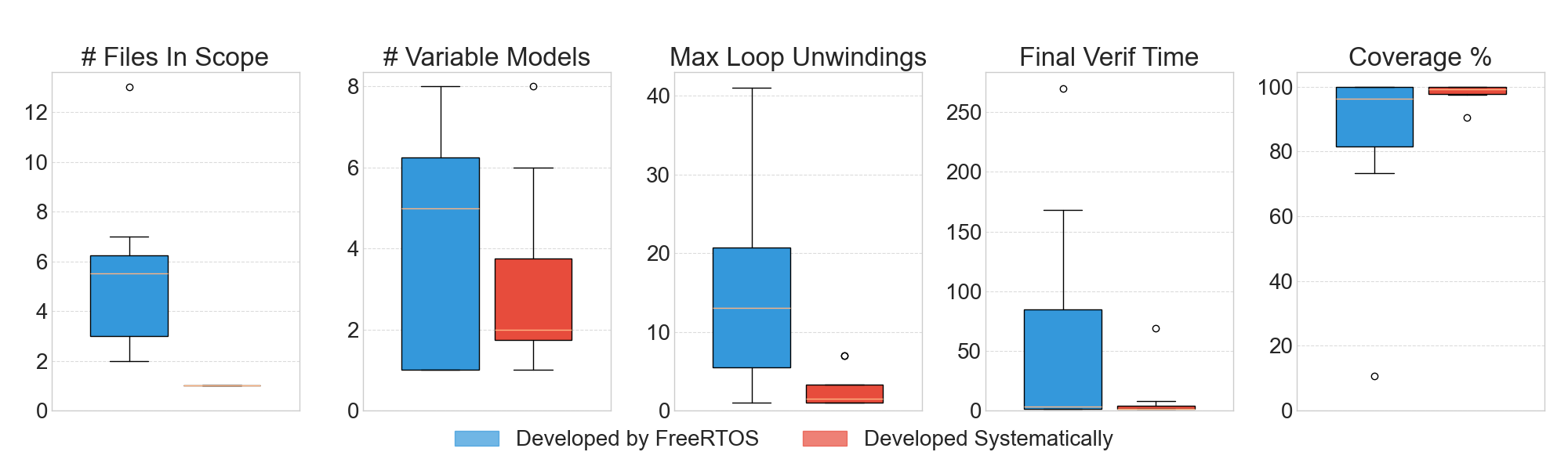}
    \caption{
    \small Comparing systematically developed FreeRTOS proofs with those developed by FreeRTOS engineers.
    }
    \label{fig:rq4-unit-proof-comparison}
\end{figure}

\Cref{fig:rq4-unit-proof-comparison} compares the unit proofs developed systematically with the existing unit proofs in FreeRTOS.
We find that the systematically developed proofs had fewer files in scope, fewer modeled variables, lower loop bounds, faster execution, and better coverage. 
Systematic proofs included additional files only when necessary to resolve violations,.
However, this preventing three defects from being detected initially (\cref{tab:rq1-vuln-detection-status}). 
In contrast, FreeRTOS proofs included all reachable files in the unit proof's scope, increasing the functional unit size and enhancing verification fidelity.
However, this also increased the number of required variables models, proof execution time and potentially, modeling effort. 

\cref{fig:rq4-unit-proof-comparison} also shows that existing FreeRTOS unit proofs achieved less coverage compared to the systematic proofs.
In one case (ProcessReceivedTCPPacket\_harness.c), the FreeRTOS proof achieved only \processReceivedTCPPracketCoverage\% coverage due to an error in the unit proof model, where an incorrect assumption fixed the packet length to an invalid value. 
This error could have been detected if coverage had been used to assess proof completeness. 
A similar issue was found in DNS\_ParseDNSReply\_harness.c, which had \ParseDNSReplyCoverage\% coverage, indicating a broader concern. 
Additional inconsistencies were also observed.
For instance, while some proofs (\eg \texttt{ProcessICMPPacket \_harness.c}) modeled packet buffers with variable lengths and an upper bound, others (\eg \texttt{DNS\_ParseDNSReply\_harness.c}) used fixed sizes. 
Prior work~\cite{amusuo_enabling_2024} has shown that using fixed sizes can prevent defects triggered by smaller packets from being detected.

\definition{
\textbf{RQ4 Finding:} The systematically developed unit proofs modeled fewer variables, executed faster and achieved better coverage. 
Observed errors and inconsistencies in existing FreeRTOS proofs indicate the need for a uniform and systematic approach.
}

\begin{figure}[h]
    \centering
    \includegraphics[width=0.99\linewidth]{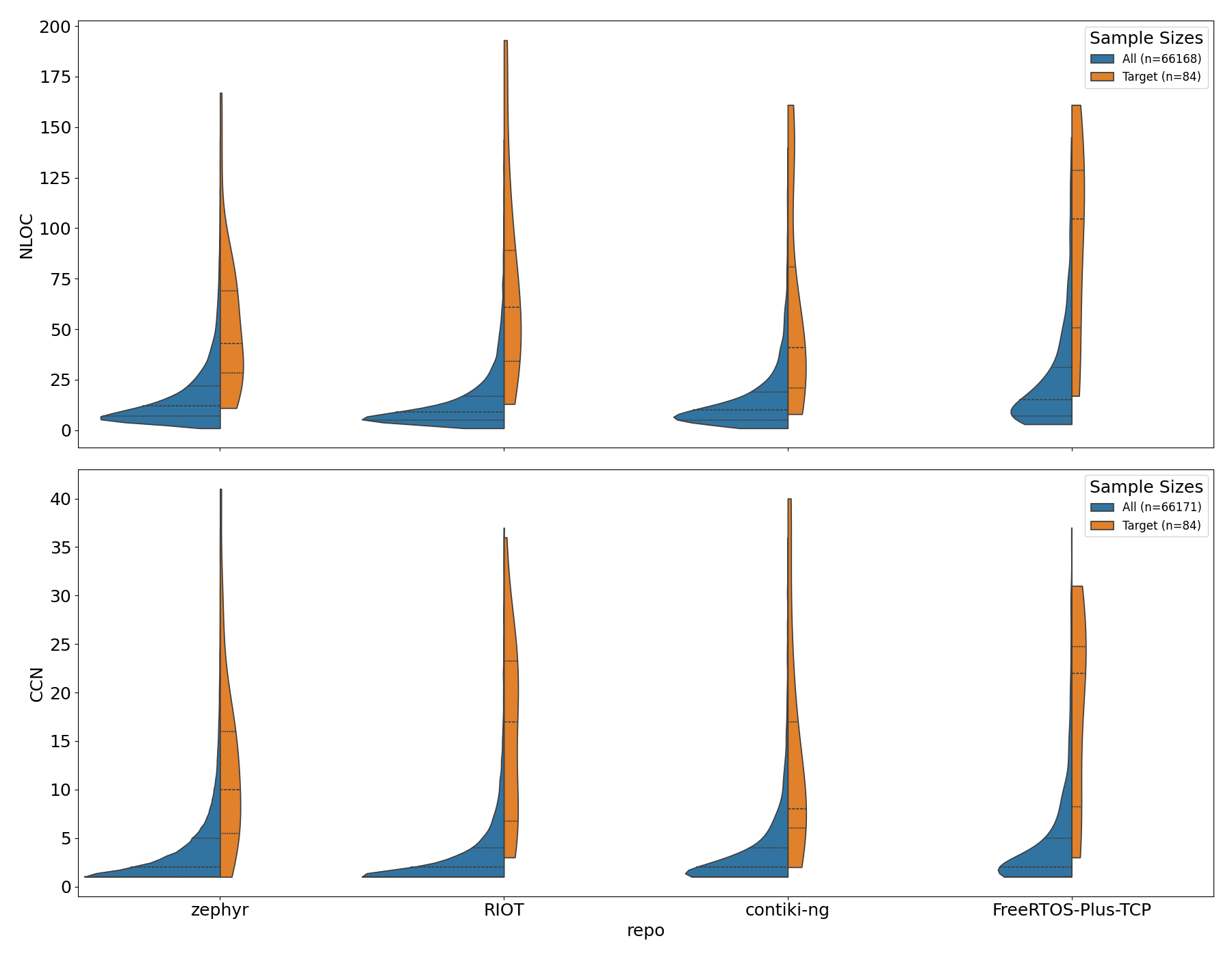}
    \caption{
    \small Comparing the size (top) and complexity (bottom) of verified functions with the functions in each embedded OS.
    CCN: Cyclomatic complexity number.
    NLOC: Number of lines of code.
    }
    \label{fig:rq5-comparisons-nloc}
\end{figure}

\subsection{RQ5: Generalizability of Results}
 
\cref{fig:rq5-comparisons-nloc} shows that the functions we developed unit proofs for are among the larger (top) and more complex (bottom) ones in each embedded OS. 
Given that unit proof size (\cref{fig:rq2-unit-proof-sizes}), development time, and execution time (\cref{subsec:rq3-results}) scale with function size, we infer that the remaining functions in each embedded OS will require similarly sized or smaller unit proofs, fewer models, and overall similar or less effort than those reported in \cref{fig:rq3-verification-step-effort}.

\definition{
\textbf{RQ5 Finding:} The developed unit proofs cover the larger and more complex functions in each embedded software, demonstrating the generalizability of our approach.
}

\section{Discussion}
\label{sec:discussion}

In this section, we discuss the broader implications of our results.

\subsection{Our Cost-benefit Analysis for Unit Proofing}

A key question asked by engineering leaders is: \textit{Should we adopt technique X?} 
This paper provides empirical data on the costs and benefits of unit proofing to support informed decisions. 
\cref{tab:rq1-vuln-detection-status} shows that following the systematic approach in \cref{subsec:method-developing-unit-proofs}, unit proofing detects up to 74\% of memory safety defects. 
With minor adjustments --- such as unrolling string-handling loops based on usage and incrementally expanding unit scope --- this increases to 83\%. 
Additionally, RQ1 results (\cref{subsec:rq1-results}) demonstrate that unit proofing supports incremental verification, enabling teams to uncover defects without verifying the entire code base. 
By writing unit proofs for \numHarnesses functions, we identified \newDefects new memory safety defects (\cref{tab:rq1-vuln-detection-status}). 
Each unit proof contributes value, either by exposing a defect or providing safety guarantees for the verified unit.  

On cost, \cref{fig:rq3-verification-step-effort} shows that the median unit proof took 72 minutes to develop and 25 seconds to execute, verifying 185 lines of C code (the \texttt{\_parse\_options()} function in RIOT~\cite{noauthor_riot_gnrc_rpl_control_messages_nodate}). 
At this rate, an engineer could verify approximately 1200 lines in one workday (8 hours).
This is significantly faster than other verification efforts, often measured in person-years~\cite{huang_lessons_2024}. 
This reduced time cost, driven by our systematic approach, requires minimal project-wide knowledge, allowing developers to write and verify unit proofs in isolation. 
For comparison, the AWS team, using their approach to unit proofing, reported that verifying 1500 lines of code took an engineer a month (or 75 lines of code per workday). 
Considering the high cost of memory safety defects~\cite{wee_here_nodate} --- which lead to 
  outages~\cite{noauthor_widespread_2024},
  security vulnerabilities~\cite{alex_rebert_safer_nodate},
  and attacks~\cite{noauthor_project_nodate}
  ---
  we conclude that unit proofing offers a cost-effective strategy for mitigating these defects during development.
Because our unit proofing approach can be applied incrementally, it need not be applied to an entire codebase, and engineers may focus on components that are particularly vulnerable (\eg user-facing) or unprotected by other means (\eg lacking hardware protections).

\subsection{Future Work: Tooling for Unit Proofing}
\label{subsec:discussion-future-work}

Our approach was conducted entirely manually.
Tooling would further reduce costs and accelerate adoption. 
Our empirical measurements can guide tool designs. 
We give three examples.

\myparagraph{(1) Modeling unknowns}
    The most time-consuming aspect of unit proofing is deriving models for unknown variables and functions (\cref{fig:rq3-verification-step-effort}), many of which could be inferred statically during the initial proof creation. 
    For instance, 55\% of derived variable models were preconditions assuming a pointer was non-null (\cref{tab:rq2-variable-models}), which could be identified through static analysis of pointer usage~\cite{ma_practical_2015, farzan_predicting_2012, xu_accelerating_2024}. 
    Similarly, 22\% of derived models constrained buffer sizes and could be inferred by analyzing indexing operations. 
    While static techniques can be imprecise, tools can be tuned to generate models only when confidence is 100\%, ensuring accuracy and improving efficiency.  

\myparagraph{(2) Loop bounds}
Tooling can also assist in deriving minimal loop bounds. 
Fixed-iteration loops require bounds based on iteration counts (\cref{subsubsec:result-rq2-characterizing-loops}), while string-handling loops should be bounded based on usage (\cref{subsec:rq1-results}). 
Together, these categories account for 59\% of loops requiring custom bounds (\cref{tab:rq2-loop-types}) and could be inferred through static analysis of loop exit conditions and buffer accesses. 

\myparagraph{(3) Analyzing error traces}
Additionally, automated tooling can analyze error traces to determine whether failures stem from unknown variables or functions and generate the necessary models to resolve these issues. 
There are algorithms for deriving preconditions
for functions~\cite{cobleigh_learning_2003, padhi_data-driven_2016, brumley_creating_2007, seghir_counterexample-guided_2013}, but they are designed for programs already in formal logic or abstract states rather than in regular programming languages. 
Adapting these algorithms to support model generation for unit proofs presents another promising yet unresolved research direction.

\vspace{0.1cm}
\noindent
Integrating these tooling advances into development environments (IDEs) would enable engineers to create unit proofs alongside software development, similar to writing unit tests.
With efficient tooling and fast execution times, engineers could verify functions in isolation before merging new code, ensuring higher reliability.
We also leave to future work, the feasibility of applying generative AI technologies such as LLMs to unit proofing.
We piloted these tools but find them currently weak, perhaps due to their probabilistic nature and the minimal real-world examples in their training data.

\section{Threats to Validity}
\label{sec:validity-threats}

\myparagraph{Construct Validity}
Multiple operationalizations of compositional bounded model checking are possible.
We specifically studied one unit proofing strategy.
Our findings may not generalize to other approaches (\cref{subsubsec:bg-unit-proofing}).

\myparagraph{Internal Validity}
\cref{subsubsec:method-rq1} discusses how we mitigated bias when creating unit proofs.
To reduce bias when characterizing unit proofs, two authors collaboratively developed the taxonomy, independently characterized the unit proofs, and then harmonized their results.
To reduce bias in reporting characteristics and cost (RQ2--3), a different author reviewed and ran all unit proofs, automatically extracting the reported quantitative data (\cref{subsec:rq3-results}). 

\myparagraph{External Validity}
Our study evaluated unit proofing on functions from embedded operating systems.
This raises concerns about generalizability. 
\textit{On what kinds of functions do our results hold?}
  RQ5 demonstrates that our study included larger, more complex functions, indicating our results represent an upper bound on unit proof size and cost in embedded software. 
\textit{Would our results hold on other embedded software?}
  Our approach was effective across the \numEmbeddedComponents diverse components in the evaluated embedded OSes, and the uniformity of functions in \cref{fig:rq5-comparisons-nloc} suggests broader generalizability to other embedded software. 
\textit{Would our results hold if other engineers applied unit proofing?}
  Part of the value of our work is that we followed a systematic approach with clear stopping criteria (see \cref{subsec:method-developing-unit-proofs} and our artifact).
  Our approach was clear enough that four authors could contribute to proof development, reducing individual bias.
  We believe that other engineers following our approach would achieve similar results.
\textit{What about other (non-embedded) software?}
  The main risk is that the constraints of embedded software lead to similar software architectures and patterns, \eg event-driven approaches and reliance on global variables.
  We cannot comment on generalizability to software designed with other constraints. 

\section{Conclusions}

As a result of many engineering disasters, academia, industry, and government have called for stronger guarantees of memory safety. 
In response, leading firms are considering a formal verification approach called unit proofing.
Despite promising experience reports, current approaches are inconsistent and detailed data on costs and benefits is missing. 
To give this discussion an empirical foundation,
  we proposed a systematic methodology for developing unit proofs and conducted the first empirical study on its effectiveness and cost. 
Following our method, we recreated $\sim$75\% of the defects we attempted, inadvertently discovered 19 new defects, and report a validation speed of around 1,200 lines of source code per day.
We conclude that unit proofing is a cost-effective strategy for ensuring memory safety. 
We highlight opportunities for tooling improvements and apply our empirical data to guide tool development.

\section{Data Availability}
\label{sec:data-availability}
Our anonymized artifact is available at: https://github.com/anon-dim98/icse26-artifacts.
It contains
  (1) developed unit proofs,
  (2) raw data and analysis scripts,
  (3) code with reintroduced defects,
  (4) details of new defects found,
  and
  (5) examples of deriving loop bounds and models from verification report.

\raggedbottom
\pagebreak
\clearpage

{\footnotesize \bibliographystyle{ACM-Reference-Format}
\bibliography{bib/main, bib/software-validation, bib/software-verification, bib/temp}}

\pagebreak

\end{document}

%% file: misc/typesetting.tex
\usepackage{algpseudocode}
\usepackage[normalem]{ulem} 
\usepackage{xspace}
\usepackage{multirow} 
\usepackage{balance} 
\usepackage{fancyvrb} 
\usepackage{caption}
\usepackage{booktabs} 

\usepackage{enumitem}
\setlist[itemize]{leftmargin=*,noitemsep,topsep=0pt}
\setlist[enumerate]{leftmargin=*}

\usepackage{etoolbox}
\makeatletter
\patchcmd{\@makecaption}
	{\scshape}
	{}
	{}
	{}
\makeatletter
\patchcmd{\@makecaption}
	{\\}
	{.\ }
	{}
	{}
\makeatother
\def\tablename{Table}

\newcommand{\eg}{\textit{e.g.,}\xspace}
\newcommand{\etal}{\textit{et al.}\xspace}

\usepackage{mdframed}
\mdfsetup{skipabove=0.5\topskip,skipbelow=0.5\topskip,align=center}

\usepackage{amsthm}
\newtheorem{thm}{Theorem}

\DeclareMathSymbol{\mlq}{\mathord}{operators}{``}
\DeclareMathSymbol{\mrq}{\mathord}{operators}{`'}

\newif\ifSAVESPACE
\SAVESPACEfalse

\ifSAVESPACE

\else

\fi

\usepackage{todonotes}
\usepackage{soul}
\usepackage{fontawesome}
\usepackage{ragged2e}

\ifDEBUG
    \newcommand{\AH}[1]{\todo[color=cyan,inline]{AH:#1}}
    \newcommand{\AM}[1]{\todo[color=red,inline]{Machiry:#1}}
    \newcommand{\JD}[1]{\todo[color=yellow,inline]{JD:#1}}
    \newcommand{\TL}[1]{\todo[color=green,inline]{SA:#1}}
    \newcommand{\PA}[1]{\todo[color=orange,inline]{PA:#1}}
    
    \newcommand{\KR}[1]{\todo[color=yellow,inline]{Kyle:#1}}
    \newcommand{\LS}[1]{\todo[color=green,inline]{LS:#1}}
    \newcommand{\HP}[1]{\todo[color=green,inline]{HP:#1}}
    \newcommand{\PP}[1]{\todo[color=lime,inline]{PP: #1}}

\else
    \newcommand{\AH}[1]{}
    \newcommand{\AM}[1]{}
    \newcommand{\JD}[1]{}
    \newcommand{\TL}[1]{}
    \newcommand{\PA}[1]{}
    \newcommand{\KR}[1]{}
    \newcommand{\LS}[1]{}
    \newcommand{\HP}[1]{}
    \newcommand{\PP}[1]{}

\fi

\newcommand{\definition}[1]{
\begin{tcolorbox} [width=1.0\linewidth, colback=blue!07!white, top=1pt, bottom=1pt, left=2pt, right=2pt]
#1
\end{tcolorbox}
}

\usepackage{hyperref}

\usepackage{cleveref}
\crefformat{section}{\S#2#1#3}
\crefname{figure}{Figure}{Figures}
\crefname{table}{Table}{Tables}
\crefname{theorem}{Theorem}{Theorems}
\crefname{thm}{Theorem}{Theorems}
\crefname{lemma}{Lemma}{Lemmata}
\crefname{equation}{Eqt.}{Eqts.}
\crefformat{Grammar}{Grammar #1}
\crefname{appendix}{Appendix}{Appendices}
\crefname{listing}{Listing}{Listings}

\newcommand{\myparagraph}[1]{\paragraph{#1}}
\renewcommand{\myparagraph}[1]{\vspace{0.25em} \noindent \hspace{0.1cm}\underline{\textit{#1:}}}

\usepackage{url}

\usepackage{tcolorbox}

\makeatletter
\newcommand{\linebreakand}{%
  \end{@IEEEauthorhalign}
  \hfill\mbox{}\par
  \mbox{}\hfill\begin{@IEEEauthorhalign}
}
\makeatother

\usepackage{pifont}

 \usepackage{pifont}
 \newcommand{\dingone}{{\color{blue}\ding{192}}}%
 \newcommand{\dingtwo}{{\color{blue}\ding{193}}}%
 \newcommand{\dingthree}{{\color{blue}\ding{194}}}%
 \newcommand{\dingfour}{{\color{blue}\ding{195}}}%
 \newcommand{\dingfive}{{\color{blue}\ding{196}}}%
 \newcommand{\dingsix}{{\color{blue}\ding{197}}}%
 \newcommand{\dingrone}{{\color{red}\ding{192}}}%
 \newcommand{\dingrtwo}{{\color{red}\ding{193}}}%
 \newcommand{\dingrthree}{{\color{red}\ding{194}}}%

%% file: misc/listing_style.tex
\usepackage{listings}
\usepackage{xcolor}

\definecolor{codegreen}{rgb}{0,0.6,0}
\definecolor{codegray}{rgb}{0.5,0.5,0.5}
\definecolor{codepurple}{rgb}{0.58,0,0.82}
\definecolor{backcolour}{rgb}{0.95,0.95,0.92}

\lstdefinestyle{mystyle}{
    backgroundcolor=\color{backcolour},   
    commentstyle=\color{codegreen},
    keywordstyle=\color{magenta},
    numberstyle=\tiny\color{codegray},
    stringstyle=\color{codepurple},
    basicstyle=\ttfamily\footnotesize,
    breakatwhitespace=false,         
    breaklines=true,                 
    captionpos=b,                    
    keepspaces=true,                 
    numbers=left,                    
    numbersep=5pt,                  
    showspaces=false,                
    showstringspaces=false,
    showtabs=false,                  
    tabsize=2
}

%% file: misc/data.tex
\newcommand{\recDefects}{89\xspace}
\newcommand{\newDefects}{19\xspace}
\newcommand{\developmentTime}{87\xspace}
\newcommand{\executionTime}{61\xspace}
\newcommand{\funcUnitrsq}{0.328\xspace}
\newcommand{\formulaSizersq}{0.852\xspace}
\newcommand{\programSizersq}{0.029\xspace}
\newcommand{\numUnitProofs}{73\xspace}
\newcommand{\numHarnesses}{73\xspace}
\newcommand{\timedUnitProofs}{21\xspace}
\newcommand{\processReceivedTCPPracketCoverage}{10.76\xspace}
\newcommand{\ParseDNSReplyCoverage}{5.75\xspace}

\newcommand{\numProtocols}{11\xspace}

\newcommand{\numEmbeddedComponents}{22\xspace}

\newcommand{\numFreeRTOSProofs}{8\xspace}

\newcommand{\proofsBelowSixtySecs}{87.67\xspace}